\documentclass[12pt]{article}
\def\be{\begin{equation}}
\def\ee{\end{equation}}
\def\bea{\begin{eqnarray}}
\def\eea{\end{eqnarray}}
\usepackage{graphicx}

\catcode`\@=11
\def\lsim{\mathrel{\mathpalette\@versim<}}
\def\gsim{\mathrel{\mathpalette\@versim>}}
\def\@versim#1#2{\vcenter{\offinterlineskip
\ialign{$\m@th#1\hfil##\hfil$\crcr#2\crcr\sim\crcr } }}
\catcode`\@=12

\catcode`@=12
\begin{document}
\thispagestyle{empty}
\begin{flushright}
UCRHEP-T508, P11044\\
July 2011\
\end{flushright}
\vspace{0.1in}
\begin{center}
{\LARGE \bf Nonzero $\theta_{13}$ for Neutrino Mixing in a\\ 
Supersymmetric $B-L$ Gauge Model\\ 
with $T_7$ Lepton Flavor Symmetry\\}
\vspace{0.5in}
{\bf Qing-Hong Cao$^{1,2,3}$, Shaaban Khalil$^{4,5}$, Ernest Ma$^6$, 
and Hiroshi Okada$^7$\\}
\vspace{0.2in}
{\sl $^1$ Department of Physics and State Key Laboratory of Nuclear Physics
and Technology, Peking University, Beijing 100871, China\\}
\vspace{0.1in}
{\sl $^2$ High Energy Physics Division, Argonne National Laboratory, Argonne, IL 60439, U.S.A\\}
\vspace{0.1in}
{\sl $^3$ Enrico Fermi Institute, University of Chicago, Chicago, Illinois 60637, U.S.A.\\}
\vspace{0.1in}
{\sl $^4$ Centre for Theoretical Physics, The British University in Egypt,\\
El Sherouk City, Postal No.~11837, P.O.~Box 43, Egypt\\}
\vspace{0.1in}
{\sl $^5$ Department of Mathematics, Ain Shams University,\\
Faculty of Science, Cairo 11566, Egypt\\}
\vspace{0.1in}
{\sl $^6$ Department of Physics and Astronomy, University of California,\\ 
Riverside, California 92521, USA\\}
\vspace{0.1in}
{\sl $^7$ School of Physics, KIAS, Seoul 130-722, Korea\\}
\end{center}
\vspace{0.2in}
\begin{abstract}\
We discuss how $\theta_{13} \neq 0$ is accommodated in a recently proposed 
renormalizable model of neutrino mixing using the non-Abelian discrete 
symmetry $T_7$ in the context of a supersymmetric extension of the Standard 
Model with gauged $U(1)_{B-L}$.  We predict a correlation between $\theta_{13}$ 
and $\theta_{23}$, as well as the effective neutrino mass $m_{ee}$ in 
neutrinoless double beta decay.
\end{abstract}

\newpage
\baselineskip 24pt

In a recent paper~\cite{ckmo11}, a supersymmetric $B-L$ gauge model with 
$T_7$ lepton flavor symmetry is proposed with the following desirable 
features. (1) Neutrino tribimaximal mixing is achieved in a renormalizable 
theory, without the addition of auxiliary symmetries and particles. (2) 
The resulting neutrino mass matrix depends on only two complex parameters, 
and is of the same form already considered some time ago~\cite{bh05}, 
using the discrete 
symmetry $A_4$~\cite{mr01,m04}. (3) The charged-lepton Yukawa sector 
exhibits a residual discrete $Z_3$ symmetry, i.e. lepton flavor 
triality~\cite{m10,cdmw11}, under which $e,\mu,\tau \sim 1,\omega^2,\omega$, 
where $\omega = \exp(2 \pi i/3) = -1/2 + i \sqrt{3}/2$.  (4) There are  
physical scalar doublets transforming as $\omega,\omega^2$ which will decay 
exclusively into leptons such that lepton flavor triality is conserved. 
(5) If the new gauge boson $Z'$ corresponding to the spontaneous symmetry 
breaking of $B-L$~\cite{hkor08} has a mass around 1 TeV, its production 
and decay into these exotic scalars may be observable at the Large Hadron 
Collider (LHC).

Recently, the T2K Collaboration has announced that a new 
measurement~\cite{t2k11} has yielded a nonzero $\theta_{13}$ for 
neutrino mixing, i.e.
\begin{equation}
0.03~(0.04) \leq \sin^2 2 \theta_{13} \leq 0.28~(0.34)
\end{equation}
for $\delta_{CP}=0$ and normal (inverted) hierarchy of neutrino masses. 
This indicates a possibly significant deviation from tribimaximal 
mixing~\cite{hps02} where $\theta_{13}=0$, $\tan^2 \theta_{12}=1/2$, and 
$\sin^2 2 \theta_{23} = 1$ are predicted.  Whereas the tribimaximal 
pattern has an elegant theoretical interpretation~\cite{m04} in terms of 
the simplest application of $A_4$~\cite{mr01}, deviations from it are 
expected~\cite{m04,mw11}.  In this paper, we present a variation of 
our previous $T_7$ proposal~\cite{ckmo11} and show how a different 
choice of the residual symmetry of the soft terms of this model will 
lead to a four-parameter neutrino mass matrix with nonzero $\theta_{13}$ and 
predicts a strong correlation between $\theta_{13}$ and $\theta_{23}$ 
as well as the effective neutrino mass $m_{ee}$ in neutrinoless double 
beta decay.

The tetrahedral group $A_4$ (12 elements) is the smallest group with a real 
\underline{3} representation. The Frobenius group $T_7$ (21 elements) is 
the smallest group with a pair of complex \underline{3} and \underline{3}$^*$ 
representations.  It is generated by
\begin{equation}
a = \pmatrix{\rho & 0 & 0 \cr 0 & \rho^2 & 0 \cr 0 & 0 & \rho^4}, ~~~ 
b = \pmatrix{0 & 1 & 0 \cr 0 & 0 & 1 \cr 1 & 0 & 0},
\end{equation}
where $\rho = \exp(2 \pi i /7)$, so that $a^7=1$, $b^3=1$, and $ab = ba^4$. 
The character table of $T_7$ (with $\xi = -1/2 + i \sqrt{7}/2$) is given by

\begin{table}[htb]
\centerline{\begin{tabular}{|c|c|c|c|c|c|c|c|}
\hline
class & $n$ & $h$ & $\chi_1$ & $\chi_{1'}$ & $\chi_{1''}$ & $\chi_3$ & 
$\chi_{3^*}$ \\
\hline 
$C_1$ & 1 & 1 & 1 & 1 & 1 & 3 & 3 \\
$C_2$ & 7 & 3 & 1 & $\omega$ & $\omega^2$ & 0 & 0 \\
$C_3$ & 7 & 3 & 1 & $\omega^2$ & $\omega$ & 0 & 0 \\
$C_4$ & 3 & 7 & 1 & 1 & 1 & $\xi$ & $\xi^*$ \\
$C_5$ & 3 & 7 & 1 & 1 & 1 & $\xi^*$ & $\xi$ \\
\hline
\end{tabular}}
\caption{Character table of $T_7$.}
\end{table}

The group multiplication rules of $T_7$ include
\begin{eqnarray}
\underline{3} \times \underline{3} &=& \underline{3}^* (23,31,12) + 
\underline{3}^* (32,13,21) + \underline{3} (33,11,22), \\  
\underline{3} \times \underline{3}^* &=& \underline{3} (2 1^*, 3 2^*, 1 3^*) + 
\underline{3}^* (1 2^*, 2 3^*, 3 1^*) + \underline{1} (1 1^* + 2 2^* + 3 3^*) 
\nonumber \\  &+& \underline{1}' (1 1^* + \omega 2 2^* + \omega^2 3 3^*) +
\underline{1}'' (1 1^* + \omega^2 2 2^* + \omega 3 3^*).
\end{eqnarray}  
Note that $\underline{3} \times \underline{3} \times \underline{3}$ has two 
invariants and $\underline{3} \times \underline{3} \times \underline{3}^*$ 
has one invariant. These serve to distinguish $T_7$ from $A_4$ and $\Delta(27)$.
We note that $T_7$ was first considered for quark and lepton masses some time 
ago~\cite{lnr07}.

Under $T_7$, let $L_i = (\nu,l)_i \sim \underline{3}$, $l^c_i \sim 
\underline{1},\underline{1}',\underline{1}'',~i=1,2,3$, $\Phi_i = 
(\phi^+,\phi^0)_i \sim \underline{3}$, 
and ${\Phi'}_i = ({\phi'}^0,-{\phi'}^-)_i \sim \underline{{3^*}}$.  
The Yukawa couplings $L_i l^c_j {\Phi'}_k$ generate the charged-lepton 
mass matrix
\bea
M_l = \pmatrix{f_1 v'_1 & f_2 v'_1 & f_3 v'_1 \cr f_1 v'_2 & \omega^2 f_2 v'_2 & 
\omega f_3 v'_2 \cr f_1 v'_3 & \omega f_2 v'_3 & \omega^2 f_3 v'_3} 
= {1 \over \sqrt{3}} \pmatrix{1 & 1 & 1 \cr 1 & \omega^2 & \omega \cr 1 
& \omega 
& \omega^2} \pmatrix{f_1 & 0 & 0 \cr 0 & f_2 & 0 \cr 0 & 0 & f_3} ~v,
\eea
if $v'_1 = v'_2 = v'_3 = v'/\sqrt{3}$, as in the original $A_4$ 
proposal~\cite{mr01}.

Let $\nu^c_i \sim \underline{{3^*}}$, then the Yukawa couplings 
$L_i \nu^c_j \Phi_k$ are allowed, with
\begin{equation}
M_D = f_D \pmatrix{0 & v_1 & 0 \cr 0 & 0 & v_2 \cr v_3 & 0 & 0} = 
{f_D v \over \sqrt{3}} \pmatrix{0 & 1 & 0 \cr 0 & 0 & 1 \cr 1 & 0 & 0},
\end{equation}
for $v_1 = v_2 = v_3 = v/\sqrt{3}$ which is necessary for consistency 
since $v'_1 = v'_2 = v'_3 = v'/\sqrt{3}$ has already been assumed for $M_l$.  
Note that $\Phi$ and ${\Phi'}$ have $B-L=0$, and both are necessary because 
of supersymmetry.

Now add the neutral Higgs singlets $\chi_i \sim \underline{3}$ and $\eta_i 
\sim \underline{{3^*}}$, both with $B-L=-2$.  Then there are two Yukawa 
invariants: $\nu^c_i \nu^c_j \chi_k$ and $\nu^c_i \nu^c_j \eta_k$ (which has 
to be symmetric in $i,j$).  Note that $\chi_i^* \sim \underline{{3^*}}$ 
is not the same as $\eta_i \sim \underline{{3^*}}$ because they have 
different $B-L$.  This means that both $B-L$ and the complexity of the 
$\underline{3}$ and $\underline{{3^*}}$ representations in $T_7$ are 
required for this scenario. The heavy Majorana mass matrix for $\nu^c$ is then
\bea
M_{\nu^c} = h \pmatrix{u_2 & 0 & 0 \cr 0 & u_3 & 0 \cr 0 & 0 & u_1} + 
h' \pmatrix{0 & u'_3 & u'_2 \cr u'_3 & 0 & u'_1 \cr u'_2 & u'_1 & 0} = 
\pmatrix{A & C & B \cr C & D & C \cr B & C & D},
\eea
where $A = h u_2$, $B = h' u'_2$, $C = h' u'_1 = h' u'_3$, and 
$D = h u_1 = h u_3$ have been assumed.  This means that the residual 
symmetry in the singlet Higgs sector is $Z_2$, whereas that in 
the doublet Higgs sector is $Z_3$.  This misalignment is different 
from that assumed previously~\cite{ckmo11}, but is nevertheless 
achievable with suitably chosen soft terms, i.e. $\chi^*_2 \chi_2$, 
${\chi'_2}^*  \chi'_2$, $\chi_2 \chi'_2$ + H.c., $\eta^*_2 \eta_2$, 
${\eta'_2}^* \eta'_2$, $\eta_2 \eta'_2$ + H.c., $\chi^*_1 \chi_1 + 
\chi^*_3 \chi_3$, ${\chi'_1}^* \chi'_1 + {\chi'_3}^* \chi'_3$, 
$\chi_1 \chi'_1 + \chi_3 \chi'_3$ + H.c., $\eta^*_1 \eta_1 + 
\eta^*_3 \eta_3$, ${\eta'_1}^* \eta'_1 + {\eta'_3}^* \eta'_3$, 
$\eta_1 \eta'_1 + \eta_3 \eta'_3$ + H.c., $\chi_2 \eta'_2$ + H.c., 
$\chi'_2 \eta_2$ + H.c., $\chi_2 (\eta'_1 + \eta'_3)$ + H.c., 
$\chi'_2 (\eta_1 + \eta_3)$ + H.c., $(\chi_1 + \chi_3) \eta'_2 $ + H.c., 
$(\chi'_1 + \chi'_3) \eta_2$ + H.c., $(\chi_1 + \chi_3) (\eta'_1 + \eta'_3)$ 
+ H.c., $(\chi'_1 + \chi'_3)(\eta_1 + \eta_3)$ + H.c.

The seesaw neutrino mass matrix is now
\bea
M_\nu = - M_D M_{\nu^c}^{-1} M_D^T 
 = {- f_D^2 v^2 \over 3~{\rm det}(M_{\nu^c})} \pmatrix{AD-B^2 
& C(B-A) & C(B-D) \cr C(B-A) & AD-C^2 & C^2-BD \cr C(B-D) & C^2-BD & D^2-C^2},
\eea
where det$(M_{\nu^c}) = A(D^2-C^2) + 2BC^2 - D(B^2+C^2)$.  Redefining the 
parameters $A,B,C,D$ to absorb the overall constant, we obtain the 
following neutrino mass matrix in the tribimaximal basis:
\begin{equation}
{\cal M}_\nu^{(1,2,3)} = \pmatrix{D(A+D-2B)/2 & C(2B-A-D)/\sqrt{2} & 
D(A-D)/2 \cr C(2B-A-D)/\sqrt{2} & AD-B^2 & C(D-A)/\sqrt{2} \cr 
D(A-D)/2 & C(D-A)/\sqrt{2} & (AD+D^2+2BD-4C^2)/2}.
\end{equation}
This is achieved by first rotating with the $3 \times 3$ unitary matrix of 
Eq.~(5), which converts it to the $(e, \mu, \tau)$ basis, then by Eq.~(10) 
below.  Note that for $D=A$ and $C=0$, this matrix becomes diagonal: 
$m_1 = A(A-B), m_2 = A^2-B^2, m_3 = A(A+B)$, which is the tribimaximal limit.  
Normal hierarchy of neutrino masses is obtained if $B \simeq A$ and inverted 
hierarchy is obtained if $B \simeq -2A$.

The neutrino mixing matrix $U$ has 4 parameters: $s_{12}, s_{23}, s_{13}$ and 
$\delta_{CP}$~\cite{pdg10}.  We choose the convention $U_{\tau 1}, U_{\tau 2}, 
U_{e3}, U_{\mu 3} \to -U_{\tau 1}, -U_{\tau 2}, 
-U_{e3}, -U_{\mu 3}$ to conform with that of the tribimaximal mixing matrix
\begin{equation}
U_{TB} = \pmatrix{\sqrt{2/3} & 1/\sqrt{3} & 0 \cr -1/\sqrt{6} & 1/\sqrt{3} & 
-1/\sqrt{2} \cr -1/\sqrt{6} & 1/\sqrt{3} & 1/\sqrt{2}}.
\end{equation}
then
\begin{equation}
{\cal M}_\nu^{(1,2,3)} = \pmatrix{m_1 & m_6 & m_4 \cr m_6 & m_2 & m_5 \cr 
m_4 & m_5 & m_3} = U^T_{TB} U \pmatrix{m'_1 & 0 & 0 \cr 0 & m'_2 & 0 \cr 
0 & 0 & m'_3} U^T U_{TB},
\end{equation}
where $m'_{1,2,3}$ are the physical neutrino masses, with
\begin{eqnarray}
m'_2 &=& \pm \sqrt{{m'_1}^2 + \Delta m^2_{21}}, \\ 
m'_3 &=& \pm \sqrt{{m'_1}^2 + \Delta m^2_{21}/2 + \Delta m_{32}^2}~~{\rm (normal
~hierarchy)}, \\
m'_3 &=& \pm \sqrt{{m'_1}^2 + \Delta m^2_{21}/2 - \Delta m_{32}^2}~~{\rm 
(inverted~hierarchy)}.
\end{eqnarray}
If $U$ is known, then all $m_{1,2,3,4,5,6}$ are functions only of $m'_1$.

\begin{figure}[htb]
\begin{center}
\includegraphics[scale=1.00]{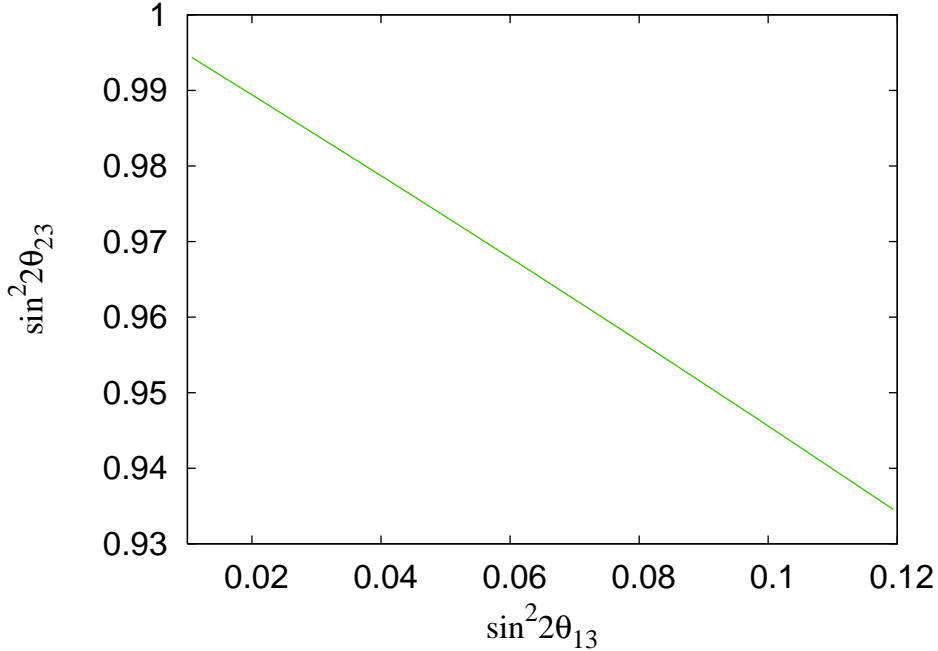}
\caption{$\sin^2 2 \theta_{23}$ versus $\sin^2 2 \theta_{13}$ for 
$\sin^2 2\theta_{12}$=0.87.\label{s12s23} }
\end{center}
\end{figure}

In our model, the neutrino mass matrix has only 4 parameters $A,B,C,D$, 
so there are 2 conditions on $m_{1,2,3,4,5,6}$.  They are given by
\begin{eqnarray}
&& A = D + {2m_4 \over D},  ~~~ B = D + {m_4 - m_1 \over D}, ~~~ 
{C \over D} = -{m_6 \over m_1 \sqrt{2}} = -{m_5 \over m_4\sqrt{2}}, \\ 
&& {D}^2 = {(m_1 - m_4)^2 \over 2m_1-m_2} = {m_1^2 (m_3 + m_1 - 2m_4) 
\over 2m_1^2-m_6^2}.
\end{eqnarray}
We now input the allowed ranges of values for $\Delta m^2_{21}$, 
$\Delta m^2_{32}$, $s_{12}, s_{23}, s_{13}$ assuming $\delta_{CP}=0$. 
In that case, $A,B,C,D$ can be chosen real. 
We then obtain $m_{1,2,3,4,5,6}$ as a function of $m'_1$.  We now  
solve for $m'_1$ using the condition $m_1 m_5 = m_4 m_6$ from Eq.(15). 
Using this value of $m'_1$, we check Eq.(16) to see if the input values 
are allowed.  In this way, we are able to find a strong correlation 
between $s_{13}$ and $s_{23}$ as shown in Fig.~\ref{s12s23}. 
It is very well approximated by
\begin{equation}
\sin^2 2 \theta_{23} \simeq 1 - {1 \over 2} \sin^2 2 \theta_{13},
\end{equation}
for all solutions. Using~\cite{pdg10} $\sin^2 2 \theta_{13} < 0.135$, 
this implies $\sin^2 2 \theta_{23} > 0.93$.

The effective neutrino mass $m_{ee}$ in neutrinoless double beta decay is 
given by
\begin{equation}
m_{ee} = {1 \over 3} |2 m_1 + m_2 + 2\sqrt{2} m_6|
= {1 \over 3} |2AD + D^2 - 2BD - B^2 + 2C(2B-A-D)|,
\end{equation}
and the kinematic $\nu_e$ mass in nuclear beta decay is $m_{\nu_e} = 
\sum_i |U^2_{ei} m'_i|$.

\begin{figure}[htb]
\begin{center}
\includegraphics[scale=0.9]{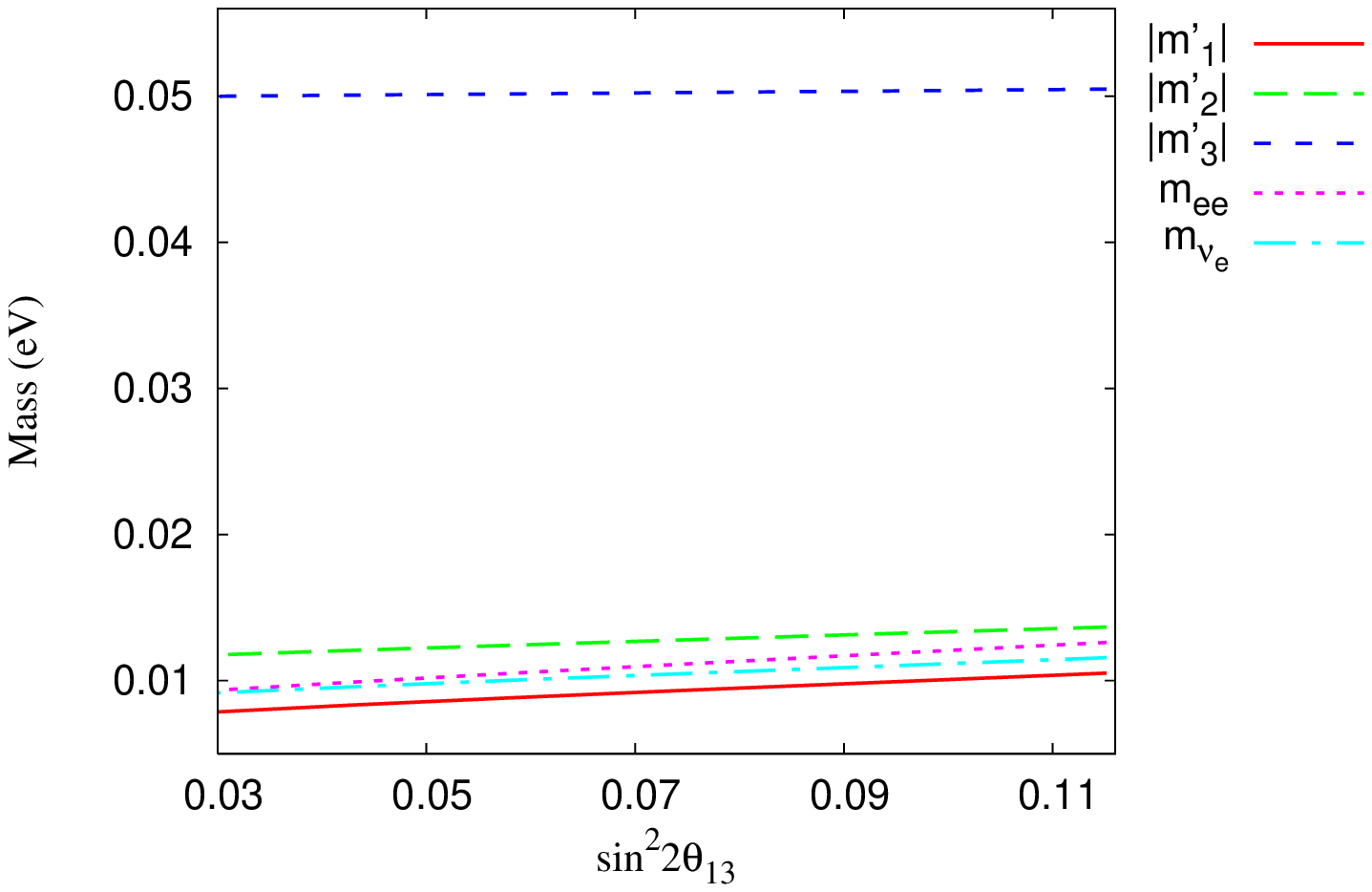}
\caption{Normal hierarchy solution in case of $\sin^2 2\theta_{12}$=0.84.
\label{NH0.84mass} }
\end{center}
\end{figure}

\begin{figure}[htb]
\begin{center}
\includegraphics[scale=0.9]{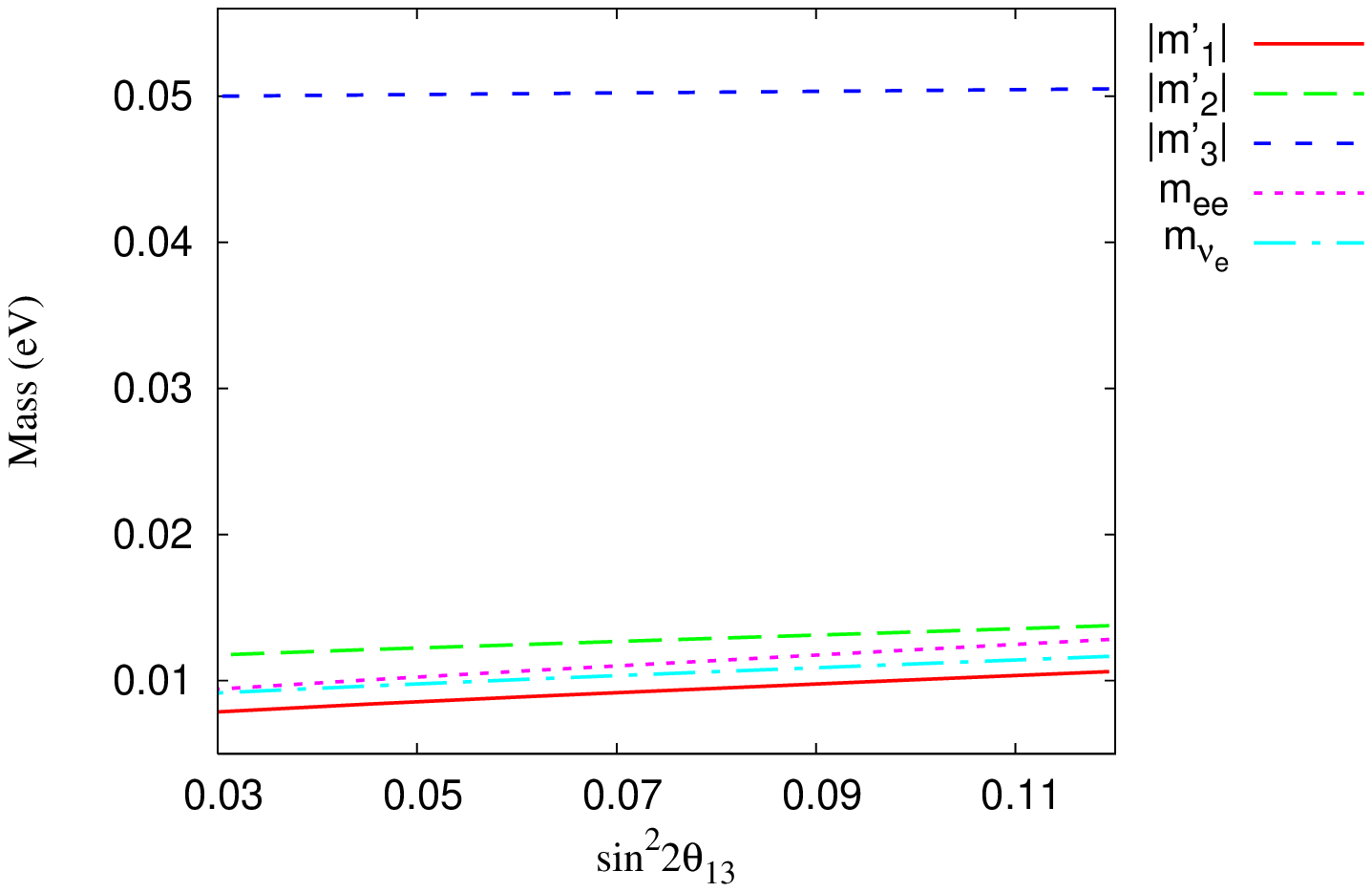}
\caption{Normal hierarchy solution in case of $\sin^2 2\theta_{12}$=0.87.
\label{NH0.87mass} }
\end{center}
\end{figure}

\begin{figure}[htb]
\begin{center}
\includegraphics[scale=0.9]{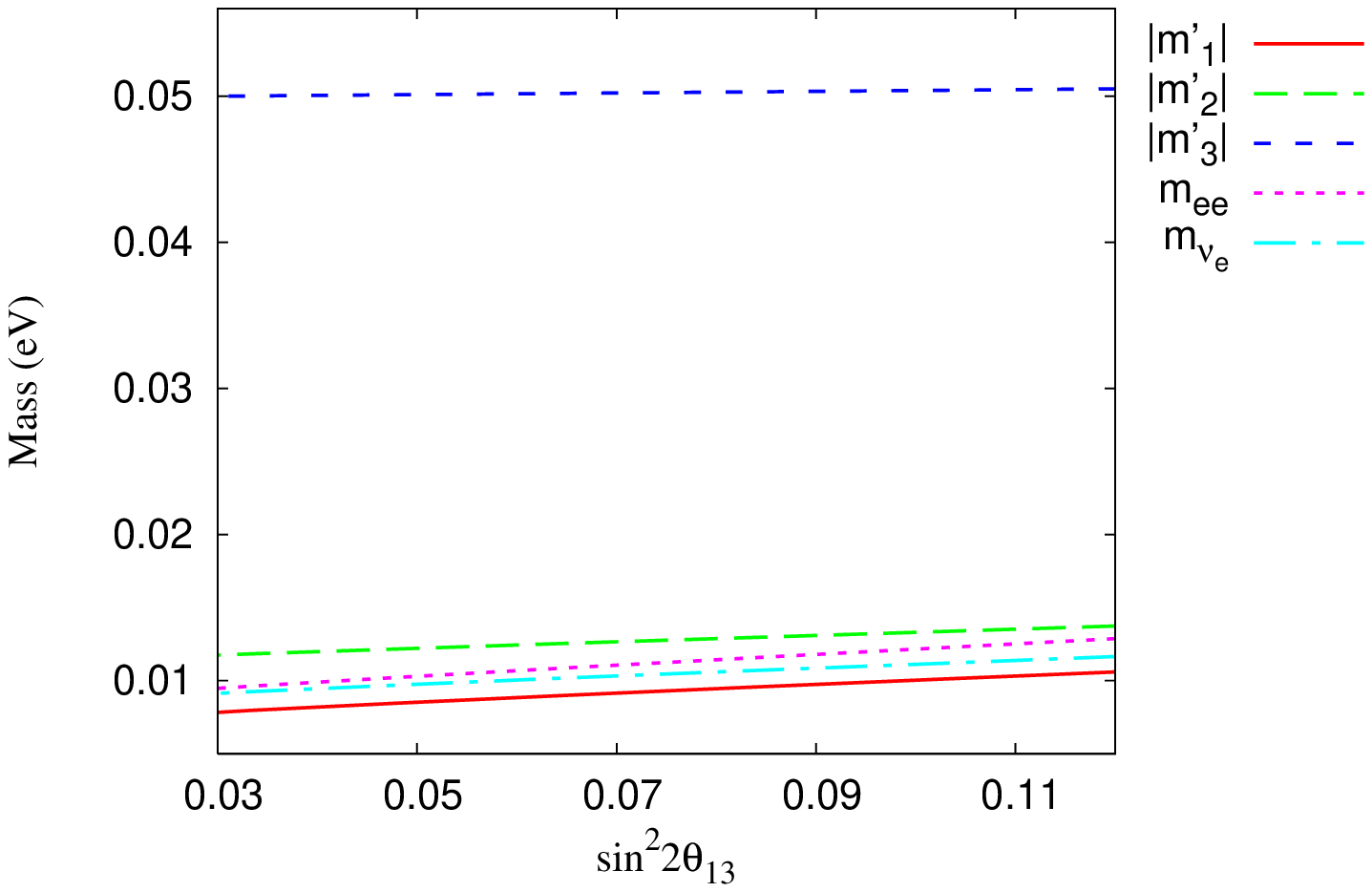}
\caption{Normal hierarchy solution in case of $\sin^2 2\theta_{12}$=0.90.
\label{NH0.90mass} }
\end{center}
\end{figure}

\begin{figure}[htb]
\begin{center}
\includegraphics[scale=0.9]{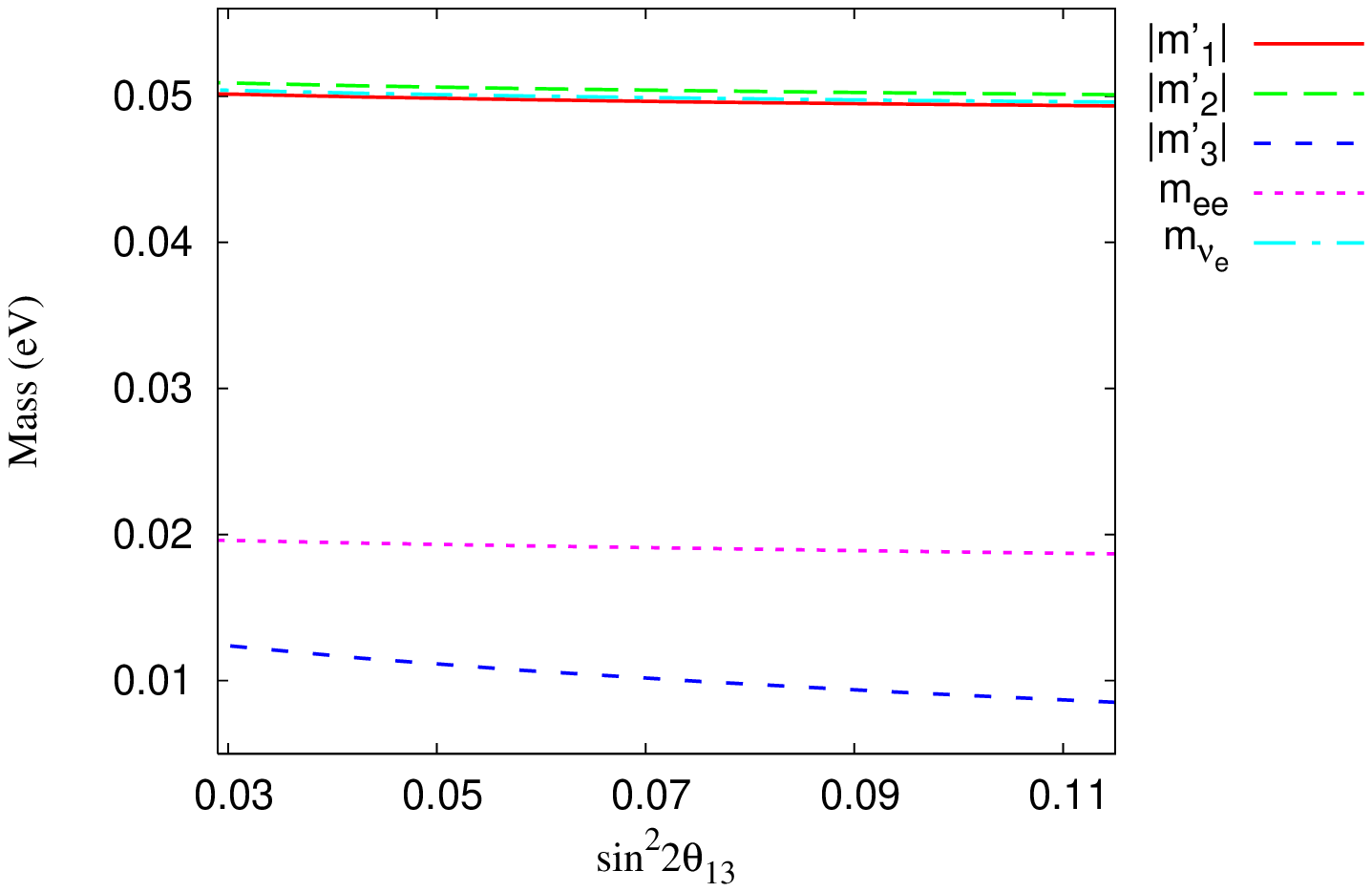}
\caption{Inverted hierarchy solution in case of $\sin^2 2\theta_{12}$=0.84.
\label{IH0.84mass} }
\end{center}
\end{figure}

\begin{figure}[htb]
\begin{center}
\includegraphics[scale=0.9]{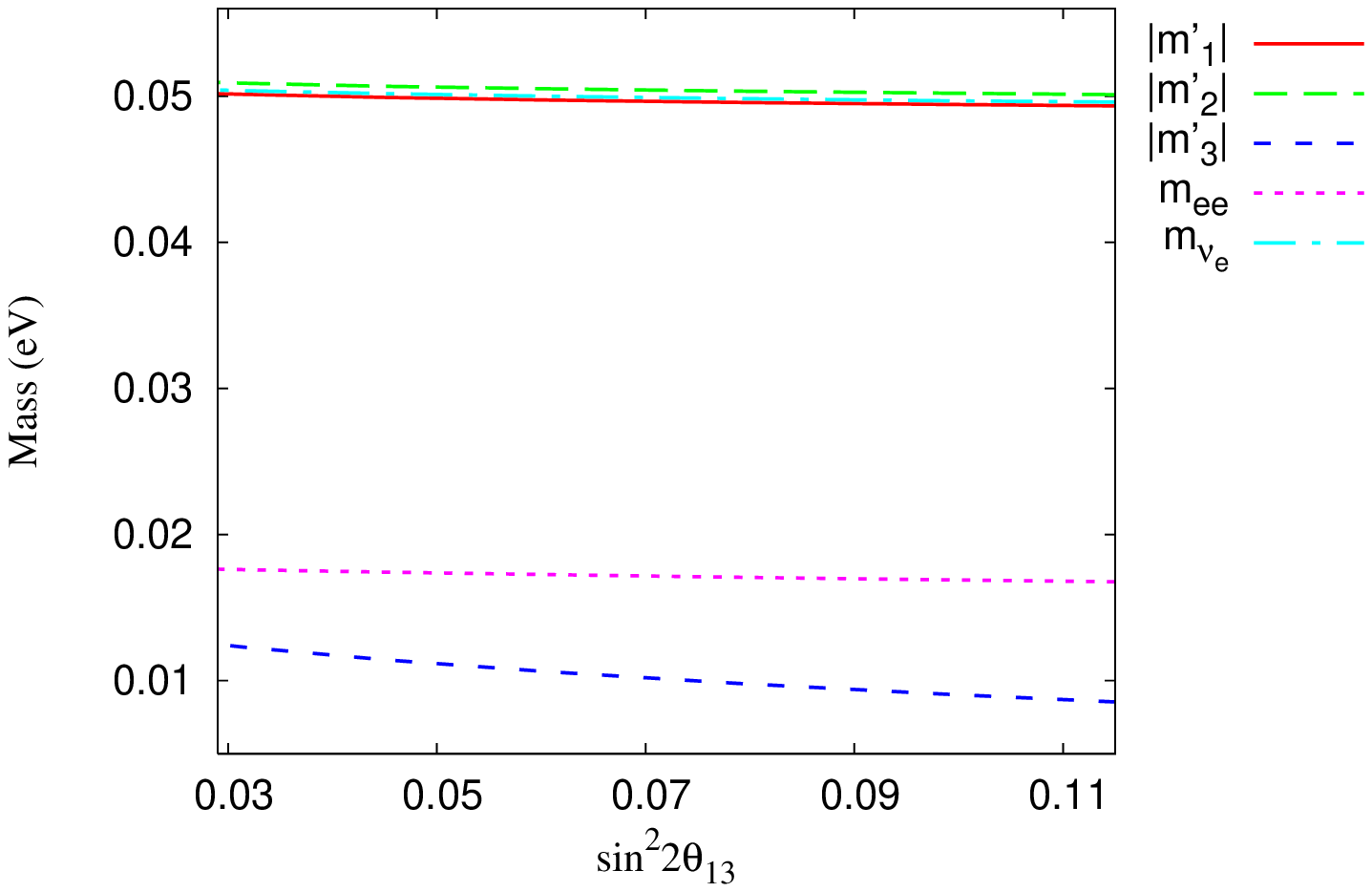}
\caption{Inverted hierarchy solution in case of $\sin^2 2\theta_{12}$=0.87.
\label{IH0.87mass} }
\end{center}
\end{figure}

\begin{figure}[htb]
\begin{center}
\includegraphics[scale=0.9]{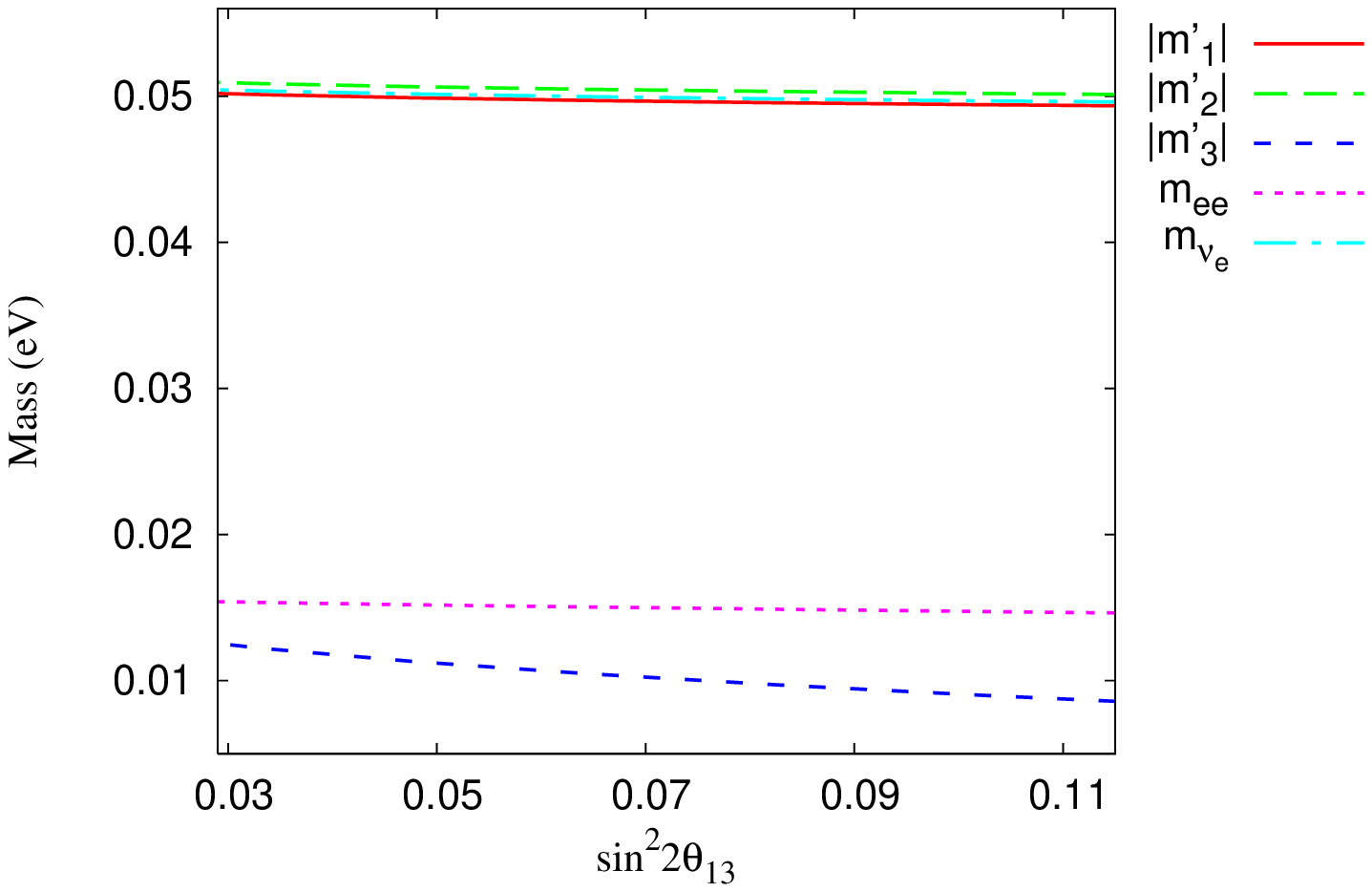}
\caption{Inverted hierarchy solution in case of $\sin^2 2\theta_{12}$=0.90.
\label{IH0.90mass} }
\end{center}
\end{figure}

\begin{figure}[htb]
\begin{center}
\includegraphics[scale=0.9]{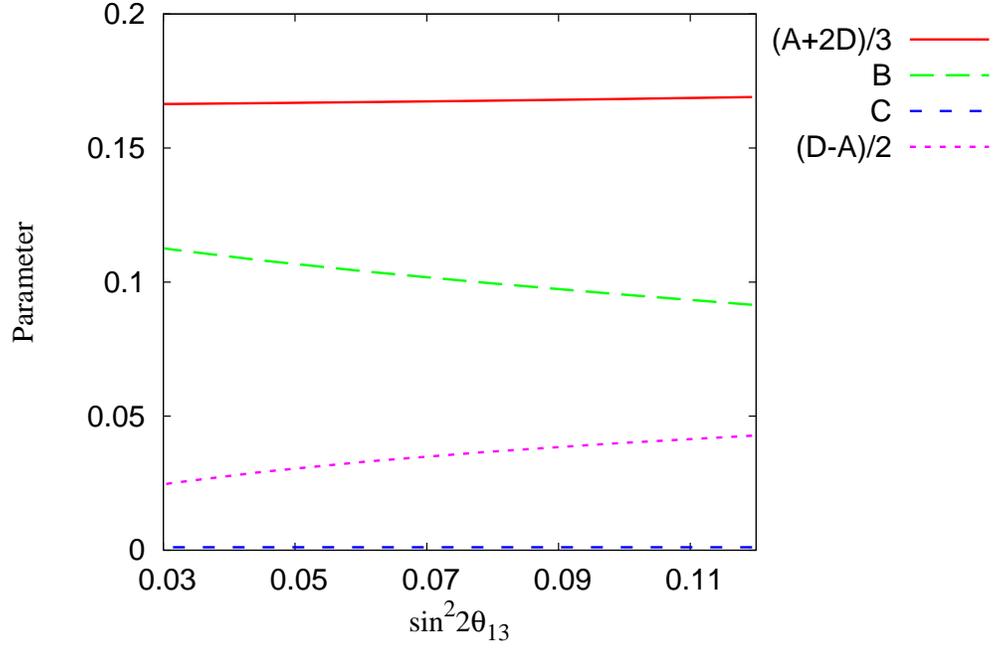}
\caption{$T_7$ parameters for normal hierarchy in case of 
$\sin^2 2\theta_{12}$=0.87.
\label{NHabcd} }
\end{center}
\end{figure}

\begin{figure}[htb]
\begin{center}
\includegraphics[scale=0.9]{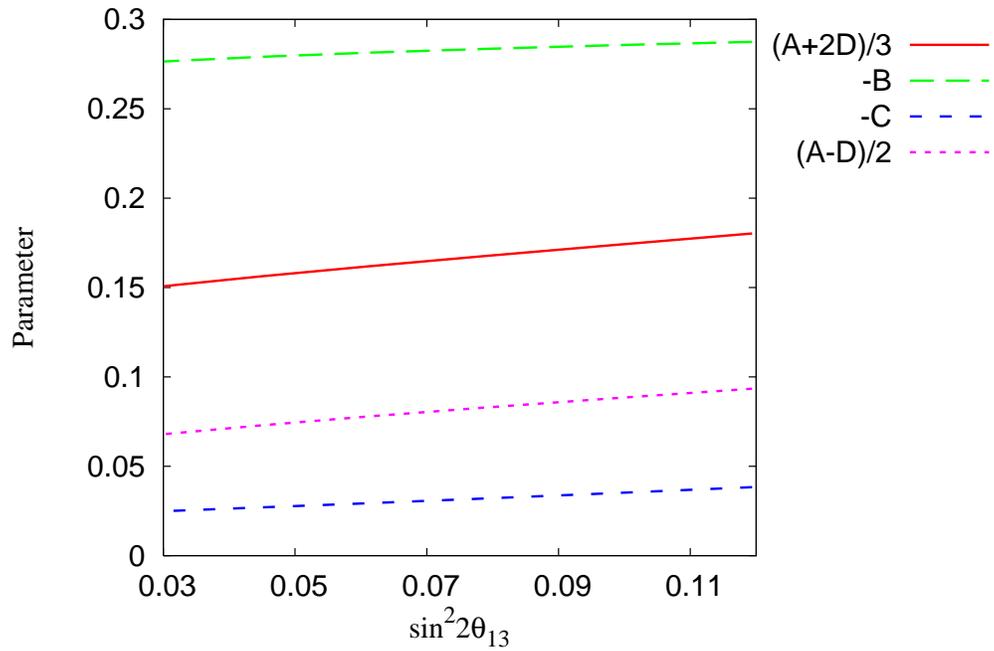}
\caption{$T_7$ parameters for inverted hierarchy in case of 
$\sin^2 2\theta_{12}$=0.87.
\label{IHabcd} }
\end{center}
\end{figure}

We find solutions for both normal and inverted hierarchies, using the 
central values of $\Delta m^2_{32} = 2.40 \times 10^{-3}$ eV$^2$ and 
$\Delta m^2_{21} = 7.65 \times 10^{-5}$ eV$^2$.  We take 3 representative 
values of $\sin^2 2 \theta_{12}$, i.e. 0.84, 0.87, 0.90.  
In Figures \ref{NH0.84mass} to \ref{NH0.90mass} we show the solutions for 
the physical neutrino masses as well as $m_{ee}$ and $m_{\nu_e}$ as functions 
of $\sin^2 2 \theta_{13}$ in the case of normal hierarchy.  In Figures 
\ref{IH0.84mass} to  \ref{IH0.90mass} we show these in the case of 
inverted hierarchy.  For $\sin^2 2 \theta_{12} = 0.87$ (corresponding to 
$\tan^2 \theta_{12} = 0.47$), we plot in Figures \ref{NHabcd} and \ref{IHabcd} 
the $T_7$ parameters $(A+2D)/3$, $B$, $C$, and $(A-D)/2$ in the case of 
normal and inverted hierarchies.
It is clear that $C$ and $(A-D)/2$ are small, 
showing that these solutions deviate only slightly from the 
tribimaximal limit.  In particular, $C=0$ exactly works for normal hierarchy, 
but it implies $\sin^2 2 \theta_{12} > 8/9$, i.e. $\tan^2 \theta_{12} > 
1/2$~\cite{m04}.

In conclusion, we have shown that a previously proposed~\cite{ckmo11} 
$T_7/B-L$ model of neutrino masses has a variation (supported by a $Z_2$ 
residual symmetry) which allows a nonzero $\theta_{13}$ and predicts 
the strong correlation $\sin^2 2 \theta_{23} \simeq 1 - \sin^2 2 \theta_{13}/2$ 
which is consistent with all data, including the recent T2K 
measurement~\cite{t2k11}.  We also predict the effective neutrino mass 
$m_{ee}$ in neutrinoless double beta decay.

The work of Q.H.C. is supported in part by the U.~S.~Dept.~of Energy  
Grant No.~DE-AC02-06CH11357 and in part 
by the Argonne National Lab.~and Univ.~of Chicago Joint Theory Institute 
Grant No.~03921-07-137. 
The work of S.K. is supported in part by the Science and Technology 
Development Fund (STDF) Project ID 437 and the ICTP Project ID 30.
The work of E.M. is supported in part by the U.~S.~Dept.~of Energy  
Grant No.~DE-FG03-94ER40837.

\newpage
\bibliographystyle{unsrt}


\end{document}